\documentclass{raa}           
\usepackage{graphicx,times}
\usepackage{natbib}
\usepackage{amssymb,amsmath}
\usepackage{color}
\usepackage[T1]{fontenc}
\usepackage{ae,aecompl}
\usepackage{soul}
\usepackage{multirow}   % plot tables
\usepackage{gensymb}    % degree symbol
\bibpunct{(}{)}{;}{a}{}{,}

\makeatletter

\newcommand{\Rmnum}[1]{\expandafter\@slowromancap\romannumeral #1@}
\makeatother

\newcommand{\HI}{\hbox{\rmfamily H\,{\textsc i}}}

\usepackage[a4paper=true,pagebackref=true]{hyperref}
\hypersetup{colorlinks = true, linkcolor = green, anchorcolor = red, citecolor = blue, filecolor = red, pagecolor = red, urlcolor = red}

\begin{document}
\title{A simulation experiment of a pipeline based on machine learning for neutral hydrogen intensity mapping surveys}
 \volnopage{ {\bf 20XX} Vol.\ {\bf X} No. {\bf XX}, 000--000}
   \setcounter{page}{1}

\author{Lin-Cheng Li \&
   Yuan-Gen Wang$^\ast$
   }
\institute{School of Computer Science and Cyber Engineering, Guangzhou University, Guangzhou 510006, China; {\it linchengli@gzhu.edu.cn, wangyg@gzhu.edu.cn}($^\ast$corresponding author)\\
\vs \no
   {\small Received 20XX Month Day; accepted 20XX Month Day}
}

\abstract{We present a simulation experiment of a pipeline based on machine learning algorithms for neutral hydrogen (\HI) intensity mapping (IM) surveys with different telescopes.
The simulation is conducted on \HI\ signals, foreground emission, thermal noise from instruments, strong radio frequency interference (sRFI), and mild RFI (mRFI).
We apply the Mini-Batch K-Means algorithm to identify sRFI, and Adam algorithm to remove foregrounds and mRFI.
Results show that there exists a threshold of the sRFI amplitudes above which the performance of our pipeline enhances greatly.
In removing foregrounds and mRFI, the performance of our pipeline is shown to have little dependence on the apertures of telescopes.
In addition, the results show that there are thresholds of the signal amplitudes from which the performance of our pipeline begins to change rapidly.
We consider all these thresholds as the edges of the signal amplitude ranges in which our pipeline can function well.
Our work, for the first time, explores the feasibility of applying machine learning algorithms in the pipeline
of IM surveys, especially for large surveys with the next-generation telescopes. 
}

\keywords{cosmology:observation, methods: statistical, radio lines: galaxy}

   \authorrunning{L. Li et al. }            %author_head in even pages
   \titlerunning{A pipeline based on machine learning}  % title_head in odd pages
   \maketitle

%________________________________________________ sections below
% 
\section{Introduction}           %% first-level sections will be auto-capitalized
\label{sect:intro}
The Large Scale Structure (LSS) traced by galaxies is a useful indicator for the information of matter distribution at high redshift, potentially allowing us to pose observational constraints on cosmological models. 
Although galaxies dim rapidly with increasing distance in the optical band, the 21-cm emission from neutral hydrogen (\HI) is luminous enough to be detectable with radio telescopes. 
Traditionally, the \HI\ signal is observed for each galaxy in the local universe for the study of galaxy formation and evolution. 
However, this observation mode is unsuitable for the intermediate redshift since the integration time required at a fixed sensitivity increases with the square of the luminous distance.
To overcome this problem, the intensity mapping (IM) survey mode has been proposed \citep{Pen08,Masui2013,CHIME2022}. 
In this mode, the total flux of galaxies instead of one galaxy in a certain direction is collected as a whole signal in one telescope beam, which saves the integration time in the survey.

With the advent of large radio telescopes, many attempts on IM have also been made with real instruments.
\citet{Pen08} reported a convincing cross-correlation signal between \HI\ Parkes All Sky Survey \citep[HIPASS,][]{Barnes01} data and optical data from the 6dF Galaxy Survey \citep[6dFGS,][]{Jones04,Jones09}.
\citet{Chang10} for the first time reported the detection of a cosmological signal at redshift $z\sim 0.8$ with IM data from the Green Bank Telescope (GBT). \citet{Masui2013} and \citet{Switzer13} followed \citet{Chang10} and further calculated the correlation function at the corresponding redshift. 
\citet{Anderson18} then studied the effect of environment on \HI\ content by cross-correlating Parkes data with the 2dFGRS strip across the South Galactic Pole.
With a phased array feed, \citet{Li2021} (hereafter Li21) detected a cross-correlation signal between optical and \HI\ density field at redshift $z\sim 0.75$.
Furthermore, \citet{CHIME2022} reported the detection of 21-cm emission from LSS between redshift 0.78 and 1.43 with the Canadian Hydrogen Intensity Mapping Experiment (CHIME).

However, there are still many challenges in the data processing in IM experiments.
The first one is the contamination from galactic and extragalactic foregrounds which are predicted to be $\sim$ 4 orders of magnitude larger than the IM temperature fluctuations.
In addition, thermal noise from observational instruments exits during the observation and is also a few orders of magnitude greater than the {\HI} signal.
Finally, radio frequency interference (RFI) prevalent at low frequencies can easily contaminate the data by many orders of magnitude. 
How to extract the weak \HI\ signal of LSS from these interferences is the key question to an IM survey.

In recent years, many studies have been made on the removal of foregrounds and extraction of \HI\ signal in IM experiments \citep{Wolz17,Cunnington2021,WJY2021,Wolz22}.
The basic idea for extracting \HI\ signal is to utilize the different properties between \HI\ signal and interference.
Thermal noise is typically uncorrelated with integration time, thereby can be suppressed by integrating over time and stacking the data within a wide range of frequencies.
The foreground emissions are mainly made of diffuse synchrotron and free-free emission, which are featureless along the frequency axis and theoretically can be removed by fitting with smoothly varying functions.
To deal with IM data and similar data from EoR experiments, researchers have also developed more sophisticated techniques in foreground removal \citep{Wang06,Liu2011,Chapman12,Switzer13,Hothi21, WJY2021}.
Generally, contamination by RFI is mitigated by flagging and removal in spectral space considering the high amplitudes of RFI.

The previous works on extracting the {\HI} signal are mainly based on traditional methods relying on manually adjusting parameters in pipelines according to results, which has limited adaptability.
An automatic pipeline based on machine learning may help improve the flexibility and independence on models. 
For example, some machine learning algorithms have been applied to solve astronomical problems, including searching for fast radio bursts \citep{Wagstaff_2016,Yang2021,Chen2022} and pulsars \citep{Eatough2010,Morello2014,Zhu2014,Wang2019,Zeng2020}, and classifications of images \citep{Aniyan2017,Lukic2018,Bastien2019}.
However, as far as we know, there has been no research on applying machine learning algorithms in the pipeline of an IM survey.
The dynamical range in which the pipeline based on machine learning can work well with different amplitudes of interference has not been investigated, either.
In this paper, we introduce an automatic pipeline based on machine learning algorithms for future large IM surveys. 
We test the performance of our pipeline on simulations of different amplitudes of interference signals, and figure out the dynamical ranges in which the pipeline functions well. 

The paper is structured as follows: Section 2 details our simulations of IM data. 
In section 3 we present our pipeline based on machine learning algorithms and show the results. In section 4 we made a summary and discussion. 
Throughout the paper, the cosmological parameters are given by \citet{Komatsu09}, with $\Omega_mh^{2}=0.1358$, $\Omega_b=0.0456$, $\Omega_\Lambda=0.726$, and $h=0.705$.

% Authors can give a citation as `\citealt{Michel+etal+1992}'.
% You may also use \cite, \citep and \citet for citation, and use Table~1
% or Figure~1 and so forth. Using \ref and \label for cross-references of
% Tables/Figures is a good way in adjusting/adding/removing text, tables or
% figures.

\section{SIMULATIONS}
To understand how our pipeline works with interference, we establish simulations that contain the signals of \HI\ density field, radio frequency interference, foreground signals and
thermal noise from observational instruments. 
Three radio telescopes with different apertures are considered in our analysis: Five-hundred-meter Aperture Spherical radio Telescope (FAST),  Parkes, and the mid-frequency instrument of Square Kilometre Array (SKA-Mid).

\subsection{Neutral hydrogen signal}
\label{sec:HI signal} 
\begin{figure}
    \centering
	\includegraphics[width=0.5\columnwidth]{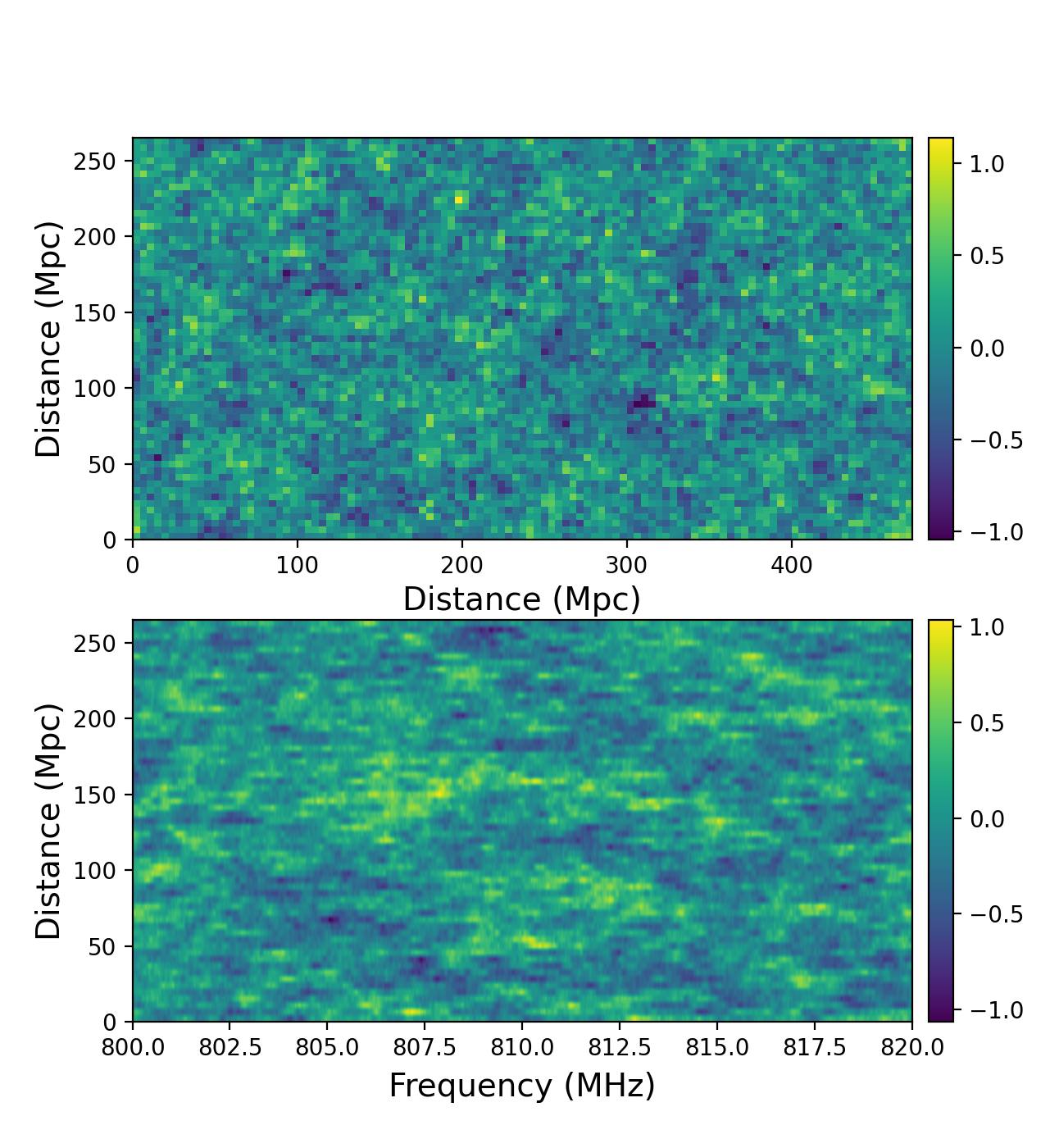}
    \caption{Two planes as examples of the simulation of \HI\ density field in the direction of line-of-sight and perpendicular to line-of-sight, respectively.}
    \label{simulation}
\end{figure}

The neutral hydrogen (\HI) signal is obtained from the simulation of 3D density fields in Li21 which details the process of generating the simulation. 
Here, we briefly explain the process: 
Firstly, the theoretical power spectrum is computed utilizing the CAMB package\footnote{https://camb.info/readme.html} \citep{Lewis00};
Then the data cube in Fourier space is generated by setting the amplitudes of moduli to corresponding power spectrum values and phases to random values; 
Finally, the data cube in the real space is computed through the inverse Fourier transformation of the data cube generated in the second step.
Figure \ref{simulation} displays two planes as examples of the simulation of \HI\ density field in the direction of line-of-sight and perpendicular to line-of-sight, respectively.

The radio beam of a telescope in our simulations is modeled with a 2D Gaussian function which width is computed with 

\begin{equation}
	\centering
    \theta \approx 1.22 \lambda/D,
	\label{del_opt}   
\end{equation}
where $\theta$ is the full width at half maximum of the Gaussian function, $\lambda$ is the electromagnetic wavelength at the observational frequency, and $D$ is the aperture of the telescope. 
Since the radio beam of a telescope has a smoothing effect on the observational data, we convolve the \HI\ density field with the Gaussian beam as the original observed \HI\ fluctuation.

In the simulations of IM, we assume that the frequency is at 810 MHz which is the central frequency of the observation in Li21. 
In addition, we set the same resolution and bandwidth of Li21 in our simulations to be in line with Li21. 
Table~\ref{keyp_simu} summarizes the key parameters of our simulations, and Table~\ref{keyp_telescope} summarizes the key parameters of telescopes considered in this paper.

\begin{table}
	\centering
	\caption{Key parameters in simulations}
	\label{keyp_simu}
	\begin{tabular}{lccr}  
		\hline
		Parameter & Value \\
		\hline
		Central frequency & 810 MHz\\
		Band width & 20 MHz \\
		Spectral resolution & 18.5 kHz\\
		Number of channels & 1080\\
		Number of cycles & 504\\
		Cycle time & 4.5 s\\
		\hline
	\end{tabular}
\end{table} 

\begin{table}
	\centering
	\caption{Key parameters of telescopes used in our work}
	\label{keyp_telescope}
	\begin{tabular}{lccr}  
		\hline
		Parameters & FAST & Parkes & SKA-Mid \\
		\hline
		Effective aperture (m) & 300 & 64 & 15\\
		beamwidth (arcmin) & 4.8 & 22.3 & 95.1\\ 
		\hline
	\end{tabular}
\end{table}

\subsection{Interference}  
In actual observations, interference signals exist in raw data. 
The patterns of interference differ with time and sites where the observations are conducted,
which makes it hard to develop a universal model for the interference.
Generally, the main interference can be categorized into three types:  thermal noise from instruments, radio frequency interference (RFI), and foreground emission.
Therefore, in our simulations we model these three types of interference and add them to the simulated data.
 
The thermal noise is mainly related to the temperature of observational instruments during the period when the observation is conducted. 
The time scale of the variations of thermal noise typically is much larger than the time span of each scan of an IM survey. 
Thus, the thermal noise in our simulations is modeled as random Gaussian noise $G\sim N(0,1)$, where $G$ is the noise matrix with the same dimensions of \HI\ data along both frequency and time axes, and $N(0,1)$ is the standard Gaussian distribution.
Before being added to the simulated data, $G$ is scaled by a scale factor ($\sigma_\textrm{t}/\sigma_\textrm{i}$) which is analysed as a varying parameter in the following sections.

The foreground signals in our simulations are made up of thermal and non-thermal radiations. 
The thermal radiation is the bremsstrahlung (free-free emission) produced by a sudden slowing down or deflection of electrons.  
In the frequency range of our simulations, bremsstrahlung has a continuous spectrum with the intensity $I(\nu)\sim \nu^{-\alpha_1}$, where $\alpha_1$ is the spectral index of the bremsstrahlung.
The non-thermal radiation is the synchrotron radiation which is generally produced in the interaction between the magnetic field and electrons. 
In the frequency range of our simulations, the synchrotron radiation also has a continuous spectrum, with the intensity $I(\nu)\sim \nu^{-\alpha_2}$, where $\alpha_2$ is the spectral index of the synchrotron radiation.
Because the frequency difference in our simulated data is very small compared to the central frequency, the summation of bremsstrahlung and synchrotron radiation in each pointing direction can be approximately expressed as a one-order polynomial (see Appendix). 
Therefore, in each cycle of the simulated data, we model the foreground signal ($F$) with
    
\begin{equation}
    F(\Delta\nu) = A\Delta\nu + B
\end{equation}
where both $A$ and $B$ are random numbers, $A\sim U(-0.5,0.5)$, $B\sim U(-0.5,0.5)$,
and $\Delta\nu$ is the spectral distance to 800 MHz in the unit of the spectral resolution multiplied by the total number of spectral channels. 
Before being added to the simulated data, the foreground signal is scaled by a scale factor ($\sigma_\textrm{HI}/\sigma_\textrm{i}$) which is analysed as a varying factor in the following analyses. 

From the IM data used in Li21, we find that the amplitudes of RFI differ broadly in different spectral channels. 
In some channels, the amplitudes of RFI can be ten times higher than the other data. 
While in some other channels, some mild RFI appears with amplitudes comparable to the foregrounds in the data.
We denote these two types of RFI as strong RFI (sRFI) and mild RFI (mRFI), respectively.
Typically, the sRFI only appears in a short range of data.
So in our simulations, sRFI is modeled to appear randomly during the 750th to 900th channels and has random amplitude that is investigated below.
Totally, two thirds of channels in that range of channels contain sRFI in the simulation. 
In spectral channels without sRFI, we add the data with mRFI which is also modeled with one-order polynomial functions with random slopes and intercepts.
In each channel that is contaminated by sRFI, the amplitude of sRFI is computed as

\begin{equation}
	\centering
    M = r_1r_2\sigma,
	\label{RFI_amplitude}   
\end{equation}
where $M$ is the mean value of the sRFI in that channel, $r_1$ is a scale factor that we investigate below, $r_2$ is a scale factor varying randomly from 0.7 to 1.3 for the uncertainties of sRFI, and $\sigma$ is the standard deviation of the data without sRFI.

\begin{figure}
    \centering
	\includegraphics[width=\columnwidth]{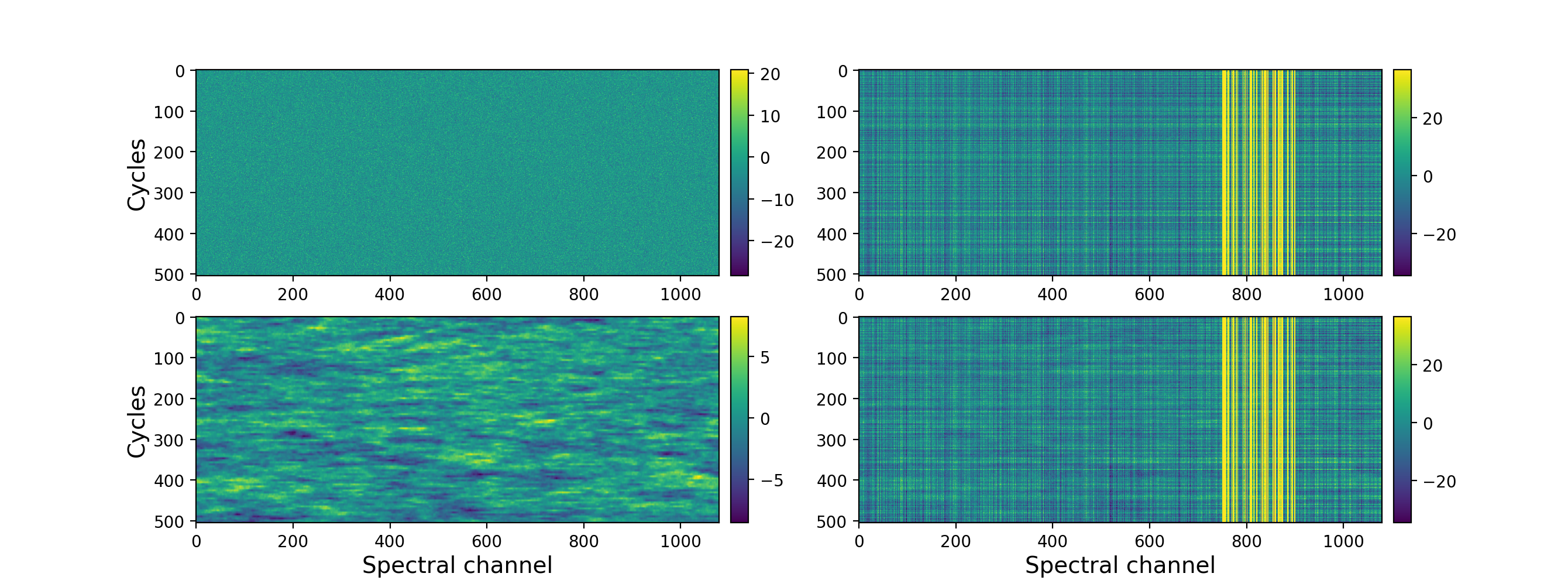}
    \caption{An example of the signal data in our simulation. 
    Top-left panel: the simulated thermal noise; 
    top-right panel: the simulated foregrounds and radio frequency interference (RFI); 
    bottom-left panel: the observed pure \HI\ signal from the simulation; 
    bottom-right panel: the overall signals from the summation of thermal noise, foregrounds, RFI, and the observed pure \HI\ signal.}
    \label{assemble}
\end{figure}

Finally, the simulated data are the summation of \HI\ signal, thermal noise, foregrounds, and RFI.
Figure \ref{assemble} shows an example of the signals in our simulation.
The simulation is used to investigate the performance of our pipeline based on machine learning algorithms.

\section{Data processing}
\subsection{sRFI identification}
Considering the high amplitudes of sRFI, data contaminated by sRFI need to be identified and separated in the first step of data processing.
We apply the Mini-Batch K-Means algorithm (MBKM) for sRFI identification given the large amount of data in our simulations and the uncertainties of the RFI amplitudes.
To better understand our pipeline, we briefly describe how the MBKM is implemented.
The MBKM is an unsupervised clustering algorithm suited well for classification in large datasets.
In our case, the raw data from simulations are classified into two classes: data contaminated by sRFI and the other data.
Firstly, the pipeline initializes two random centers as the mean values of the two classes to be classified.
Then in each epoch, the pipeline randomly extracts a part of data from the simulation and classifies data into contaminated and the other data based on the difference between the centers and the data values. 
Then new centers are updated by averaging on the classified data.
This process is repeated until the classification result converges.

\begin{figure}
    \centering
	\includegraphics[width=\columnwidth]{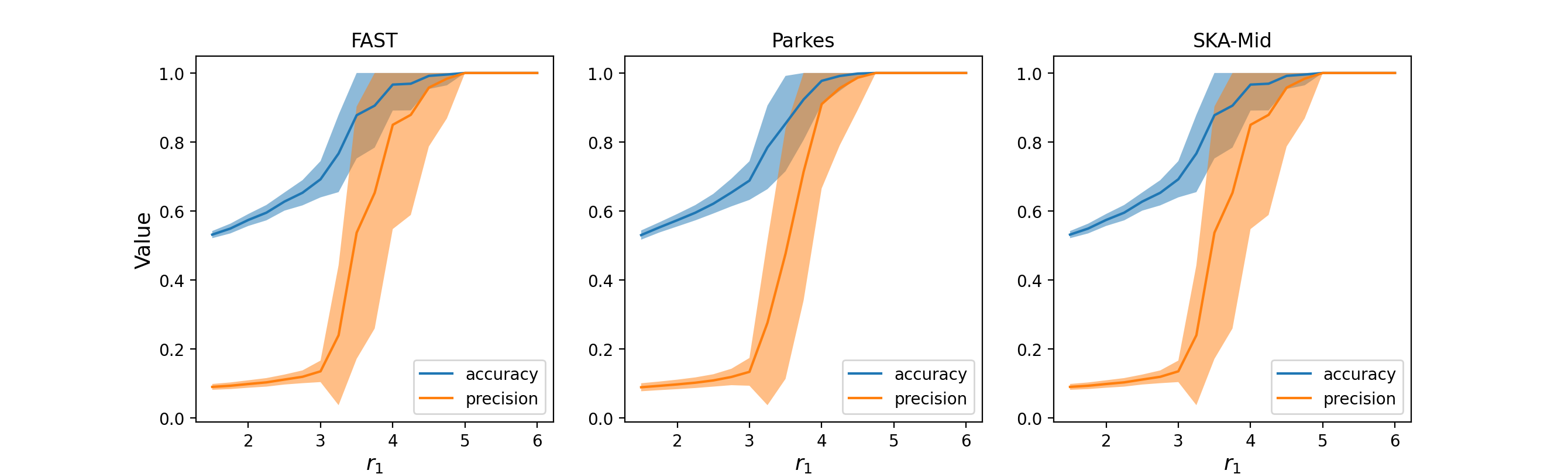}
    \caption{The accuracy and precision values in our pipeline as functions of mean amplitudes of sRFI for different telescopes.
    Blue and orange shades indicate the range from the 25th to the 75th percentiles over 50 simulations for $P$ and $A$, respectively.}
    \label{MBKM}
\end{figure}

To evaluate the performance of our pipeline in the process of sRFI identification, we compute the accuracy ($A$) and precision ($P$) with

\begin{equation}
    A = N_{\rm{c}}/N_{\rm{t}},
\end{equation}
and
\begin{equation}
    P = N_{\rm{cR}}/N_{\rm{tR}},
\end{equation}
where $N_{\rm{c}}$ is the number of data that are correctly classified, $N_{\rm{t}}$ is the total number of data, $N_{\rm{cR}}$ is the number of contaminated data that are correctly classified, and $N_{\rm{tR}}$ is the total number of contaminated data classified by MBKM.
We plot $A$ and $P$ as functions of the mean amplitudes of sRFI ($r_1$) in Figure \ref{MBKM} with different telescopes. 
As shown in Figure \ref{MBKM}, both $A$ and $P$ for each telescope increase rapidly when $r_1$ increases from 3 to 5, indicating the threshold of $r_1\sim4$ above which our pipeline is able to identify the sRFI. 
When $r_1$ is larger than 5, all the sRFI can be identified correctly in our pipeline.
In real observations, the threshold of 5 can be reached easily given that the amplitude of sRFI is typically 10 times higher than the other data.
Therefore, in the following analysis, we assume that all sRFI has been correctly identified and those data contaminated by sRFI have been discarded.

\subsection{Removal of foregrounds and mRFI}
After identifying and discarding data contaminated by sRFI, the foregrounds and mRFI are removed together in this step since their amplitudes are at comparable levels.
The algorithm that we applied to the simulation in this step is the Adaptive Momentum Estimation (Adam), which is a prevalent gradient descent algorithm in the machine learning field. 
We briefly explain how it works here.
In the Adam algorithm, the targeted loss is computed as 

\begin{equation}
	\centering
    L = ||M-S||_2 = \sum_{i}(M_i-S_i)^2,
\end{equation}
where $L$ is the loss between $S$ and $M$, $S$ is the simulated data in our mock observation, and $M$ is the fitting matrix of foregrounds and mRFI.
Our goal is to make $L$ as small as possible.
Therefore, in each cycle and each frequency channel of $M$, we fit polynomial functions to foregrounds and mRFI in $S$.

\begin{figure}
    \centering
	\includegraphics[width=\columnwidth]{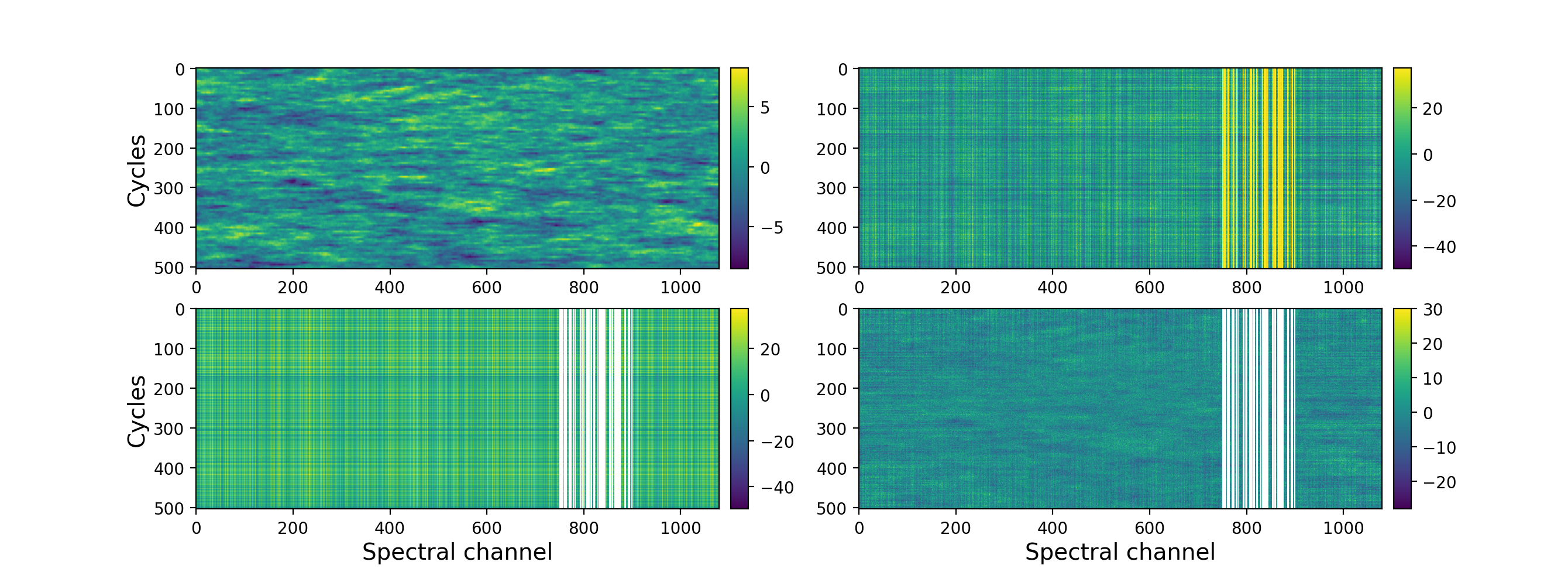}
    \caption{An example of the result of extracting the observed \HI\ density field from the noise, foreground signals, and RFI. 
    Top-left panel: the observed pure \HI\ signal from the simulation; 
    top-right panel: the overall signal data generated from the summation of thermal noise, foregrounds, RFI, and the observed pure \HI\ signal; 
    bottom-left panel: the fitting foregrounds and mRFI after applying MBKM and Adam algorithm; 
    bottom-right panel: the \HI\ signal extracted from the overall signal. 
    White channels are masked due to sRFI contamination.}
    \label{extraction}
\end{figure}

As a result, $L$ is related to the parameters $W$ used in calculating $M$. 
By computing the gradient of $L$ over $W$ ($\Delta W$), we can adjust $W$ and get a smaller $L$ and thereby a better $M$. 
To make the iterative process converge more quickly, Adam algorithm uses the moving average method to adjust the step for which the process of gradient descent takes to move forward, which can be expressed as follows:

\begin{equation}
	\centering
    V_{(t)} = \beta_1V_{(t-1)} + (1-\beta_1)\Delta W_{(t)i},
\end{equation}
\begin{equation}
	\centering
    S_{(t)} = \beta_2S_{(t-1)} + (1-\beta_2)\Delta W_{(t)i}^2,
\end{equation}
and
\begin{equation}
	\centering
    W_{(t)i} = W_{(t-1)i} - \frac{\eta}{\sqrt{S_{(t)}}+\epsilon}V{(t)},
\end{equation}
where typically $\beta_1=0.9$ and $\beta_2=0.999$ are hyper-parameters, $\eta$ is the moving length for each step, and $\epsilon=10^{-7}$ is a tiny value to keep Adam stable. 
If we keep performing the process described above, $L$ can finally reach a small enough value, which indicates that the foregrounds and mRFI have been well fitted.
So the \HI\ signal can be extracted by subtracting the fitting signals from $S$.
Figure \ref{extraction} shows an example of the process of extracting the \HI\ density field from the thermal noise, foregrounds and RFI.

\subsubsection{Dependence on the orders of polynomials in fitting}
\begin{figure}
    \centering
	\includegraphics[width=0.6\columnwidth]{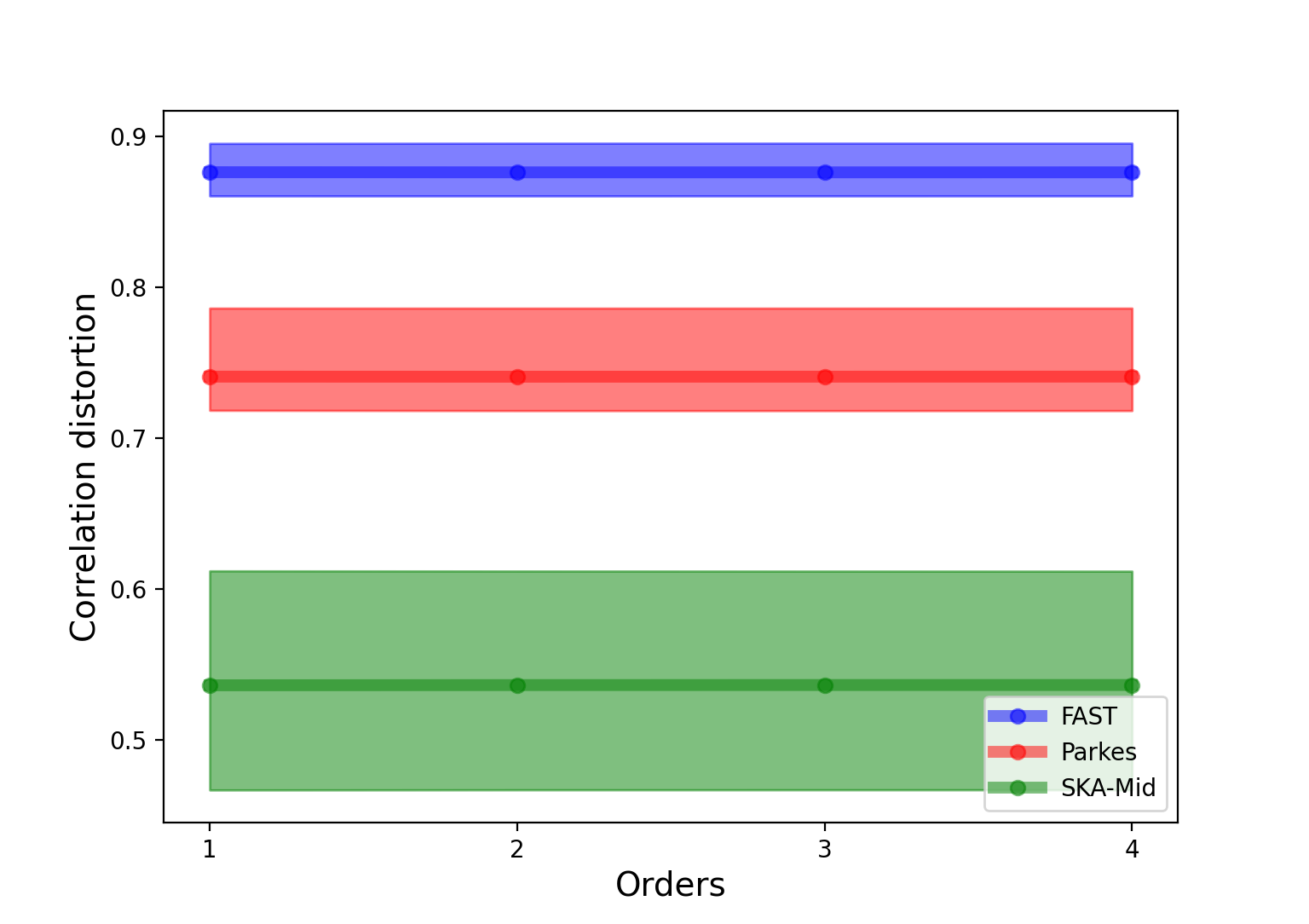}
    \caption{The median values of correlation distortion with different orders of the polynomials used in the fitting process over 50 simulations with different telescopes. Shades with corresponding colors indicate the range between 25th and the 75th percentiles over the 30 simulations.
    }
    \label{order}
\end{figure}

Theoretically, the foreground signals mainly consist of free-free emission, which power spectrum has a power-law shape along the frequency axis. 
Thus, the foreground signals can be fitted well with a one-order polynomial.
To investigate whether higher orders of polynomial functions can help improve the final result, we test different orders of polynomial functions used in the fitting process of our pipeline. 
During the test, $\sigma_\textrm{t}/\sigma_\textrm{i}$ is set to 1 and $\sigma_\textrm{HI}/\sigma_\textrm{i}$ in our simulations is set to 0.1, which is at the center of the parameter range that we consider in the analysis below.
We compare the results of correlation distortion with different orders of polynomial functions used in fitting.
The results are shown in Figure \ref{order}.
We can see little relationship between the correlation distortion and the orders of polynomial functions.
Considering that higher orders of polynomial functions result in more fitting parameters and thereby require more computational time, we use one-order polynomials as the fitting function in the following analysis. 

\subsubsection{Dependence on the amplitudes of interference}
\begin{figure}
    \centering
	\includegraphics[width=\columnwidth]{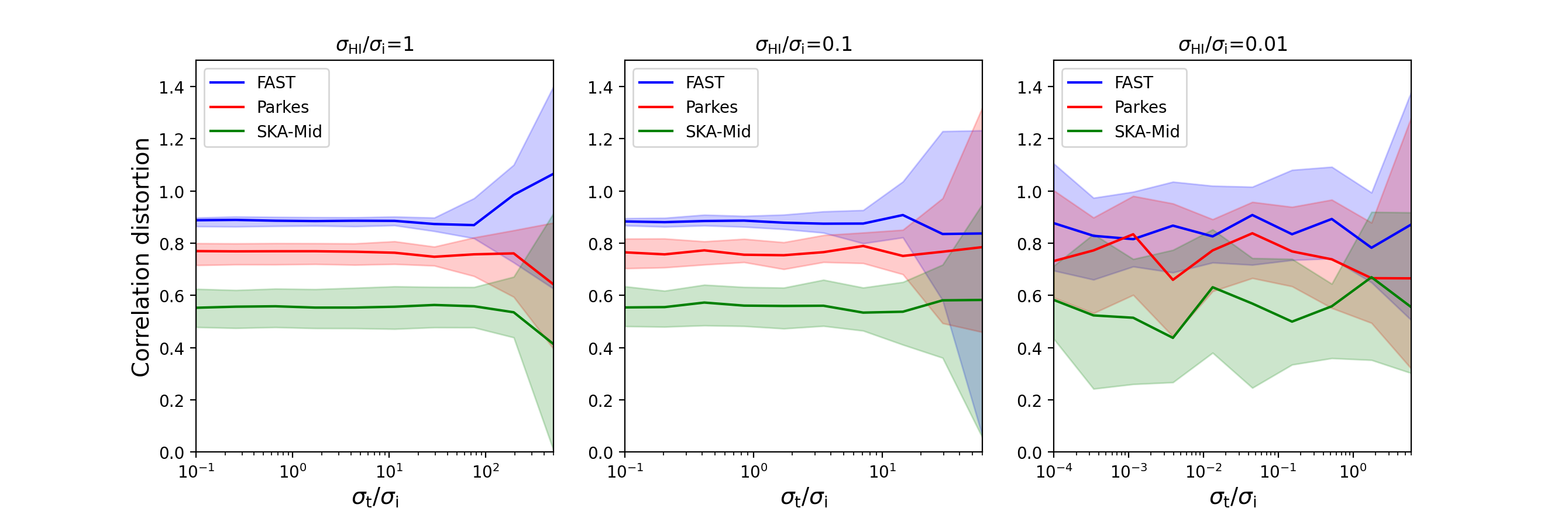}
    \caption{The correlation distortion as functions of $\sigma_\textrm{t}/\sigma_\textrm{i}$ with different telescopes. 
    Shades indicate the range from the 25th to the 75th percentiles over 50 simulations.
    Different colors indicate the results for different telescopes.
    Left to right panels are results with $\sigma_\textrm{HI}/\sigma_\textrm{i}$=1, 0.1 and 0.01, respectively.
    %The vertical dash line in each panel indicate the threshold above which the shades enlarge rapidly with increasing $\sigma_\textrm{HI}/\sigma_\textrm{i}$.
    }
    \label{ratio1}
\end{figure}

In order to assess the performance of our pipeline and figure out the dynamical range in which our pipeline works in data containing different amplitudes of interference, we compute the correlation distortion ($T$) with

\begin{equation}
	\centering
    T = C_{\rm{c}}/A_{\rm{c}},
\end{equation}
where $C_{\rm{c}}$ is the cross-correlation value between the extracted \HI\ density fields and original \HI\ density fields, and $A_{\rm{c}}$ is the auto-correlation value of the original \HI\ density fields.
The investigation is conducted with changes of $\sigma_\textrm{t}/\sigma_\textrm{i}$ and $\sigma_\textrm{HI}/\sigma_\textrm{i}$, where $\sigma_{t}$ is the standard deviation of thermal noise, $\sigma_{i}$ is the standard deviation of the summation of foregrounds and mRFI, and $\sigma_\textrm{HI}$ is the standard deviation of \HI\ density fields.
Figure \ref{ratio1} shows the results. 
From Figure \ref{ratio1} we can see a strong connection between the correlation distortion and the apertures of telescopes.
Specifically, a telescope with a larger aperture has a higher value of $T$, demonstrating the better ability of telescopes with greater apertures to resolve the observed fields. 
Interestingly, in each panel there is a threshold of  $\sigma_\textrm{t}/\sigma_\textrm{i}$ above which the errors of $T$ begin to increase rapidly with increasing $\sigma_\textrm{t}/\sigma_\textrm{i}$ for all telescopes. 
We treat this threshold of $\sigma_\textrm{t}/\sigma_\textrm{i}$ as the upper limit below which our pipeline can function well. 
In addition, the threshold of $\sigma_\textrm{t}/\sigma_\textrm{i}$ shows little relationship with the apertures of telescopes, but is dependent on $\sigma_\textrm{HI}/\sigma_\textrm{i}$, decreasing from 10 to 0.2 with $\sigma_\textrm{HI}/\sigma_\textrm{i}$ decreasing from 1 to 0.01. 

\begin{figure}
    \centering
 	\includegraphics[width=\columnwidth]{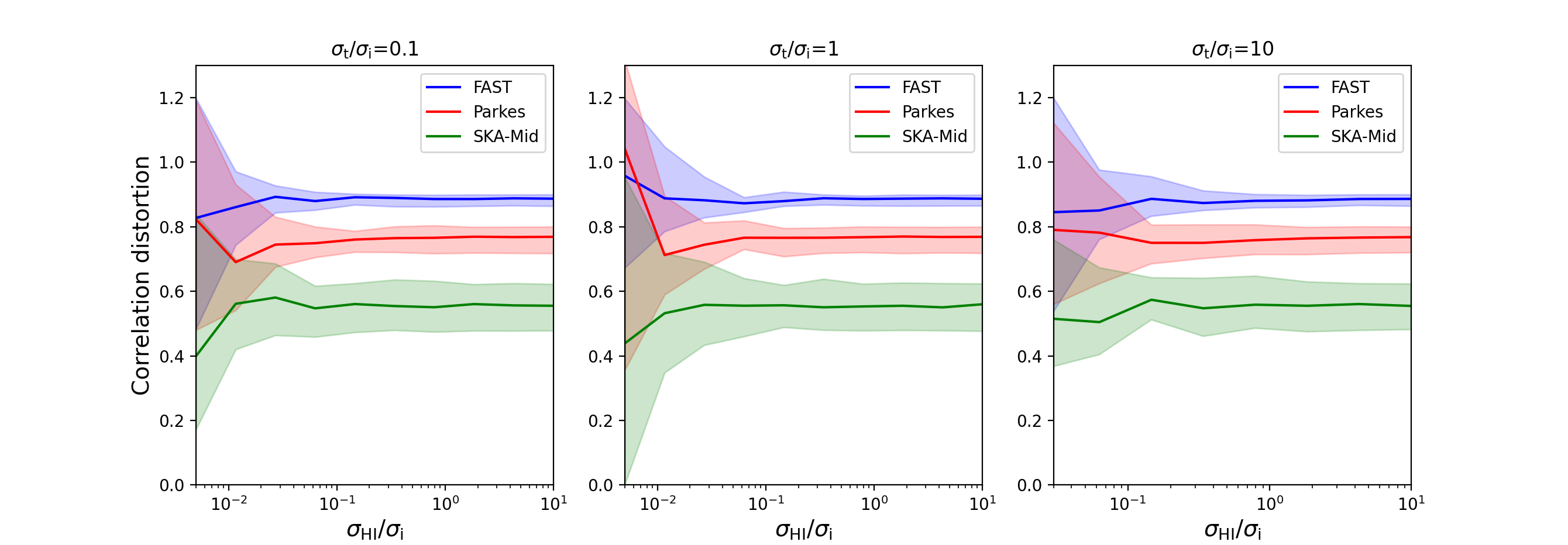}
    \caption{The correlation distortion as functions of $\sigma_\textrm{HI}/\sigma_\textrm{i}$ with different telescopes. 
    Shades indicate the ranges between the 25th and the 75th percentiles over 50 simulations.
    Different colors indicate the results for different telescopes. 
    Left to right panels are results with $\sigma_\textrm{t}/\sigma_\textrm{i}$=0.1, 1 and 10, respectively.
    %The vertical dash line in each panel indicate the threshold below which the shades enlarge rapidly with decreasing $\sigma_\textrm{HI}/\sigma_\textrm{i}$.
    }
    \label{ratio2}
\end{figure}

In Figure \ref{ratio2}, we fix the value of $\sigma_\textrm{t}/\sigma_\textrm{i}$ and vary $\sigma_\textrm{HI}/\sigma_\textrm{i}$ to investigate the dependence of the amplitude of \HI\ signal.
The results in Figure \ref{ratio2} show that in each panel there is also a threshold of $\sigma_\textrm{HI}/\sigma_\textrm{i}$ below which the errors of correlation distortion begin to increase remarkably.
We consider this threshold of $\sigma_\textrm{HI}/\sigma_\textrm{i}$ as the lower limit above which our pipeline can work well.
Furthermore, this threshold of $\sigma_\textrm{HI}/\sigma_\textrm{i}$ is dependent on $\sigma_\textrm{t}/\sigma_\textrm{i}$, rising from 0.1 to 1 with $\sigma_\textrm{t}/\sigma_\textrm{i}$ increasing from 0.1 to 10.
The threshold of $\sigma_\textrm{HI}/\sigma_\textrm{i}$ also shows little dependence on the aperture of the telescope, which confirms the result in Figure \ref{ratio1}. 

\section{Summary and conclusion}
In this paper, we have conducted a simulation experiment of the pipeline for IM surveys with different telescopes, which is the first time that the machine learning algorithms are applied in the IM pipeline.    
The simulations that we developed are conducted on \HI\ signal, foregrounds, thermal noise from instruments, and RFI.
During the process of identifying sRFI, we applied the MKBM algorithm to identify the data contaminated by sRFI. 
The performance of our pipeline in this process has also been investigated by computing the accuracy and precision values. 
The result shows that there exists a threshold of $r_1$ at which the accuracy and precision values change dramatically. 
When $r_1$ is above 5, our pipeline is able to correctly identify all the data contaminated by sRFI.

In the removal of foregrounds and mRFI, we have applied Adam algorithm to the data with different amplitudes of interference.
We have compared the results from fitting with different orders of polynomial functions, but found little dependence on the orders of polynomials.
To investigate the performance of our pipeline and figure out the dynamical range in which our pipeline can work, we calculated the correlation distortion and changed the value of $\sigma_\textrm{t}/\sigma_\textrm{i}$ and $\sigma_\textrm{HI}/\sigma_\textrm{i}$ independently. 
The results show that the performance of our pipeline has little dependence on the apertures of telescopes.
In addition, we found thresholds of $\sigma_\textrm{t}/\sigma_\textrm{i}$ and $\sigma_\textrm{HI}/\sigma_\textrm{i}$ at which the performance of our pipeline begins to change rapidly, which we consider as the indicators of the ranges in which our pipeline can function well.
As expected, the thresholds change with $\sigma_\textrm{t}/\sigma_\textrm{i}$ and $\sigma_\textrm{t}/\sigma_\textrm{i}$, showing dependence on the amplitudes of signals in an IM survey.

On the other hand, we would like to remind the reader that our work is based on the simulated observation within a relatively short frequency range (800$\sim$820 MHz) so that we can simplify the foreground model. 
If the data frequency range is wide enough in another observation, the error caused by the approximation made in the Appendix may not be negligible anymore. 
Then whether higher orders of polynomials can result in better results in the process of foreground removal needs to be investigated again.

Overall, our work has demonstrated the feasibility of processing the raw data from an IM survey based on machine learning algorithms. 
The results in the paper are encouraging and may have shed light on the potential of applying more powerful machine learning algorithms in future \HI\ IM surveys with the next-generation telescopes.
 
\normalem
\begin{acknowledgements}
We thank anonymous reviewers for their helpful comments and suggestions on improving the paper. 
This work was supported by the National Natural Science Foundation of China under Grants 61872099 and 62272116.
\end{acknowledgements}
  
\bibliographystyle{raa}
\bibliography{ms2022-0207}
 
\appendix
\section{Foregrounds with different spectral indices}
In the simulation, the foreground signals consist of synchrotron and bremsstrahlung (free-free emission).
The radiation intensity of synchrotron or bremsstrahlung in the frequency range that we consider in this paper can be expressed as 

\begin{equation}
    I(\nu) = A\nu^{-\alpha} + B
    \label{intensity}
\end{equation}
where both $A$ and $B$ are coefficients, $\nu$ is the frequency, and $\alpha$ is the spectral index.
In our case, the frequency difference $\Delta \nu \ll 800$ MHz.  
Equation \ref{intensity} thereby can be expressed with Taylor polynomial as

\begin{equation}
\begin{aligned}
I(\nu) &= I(\nu_0) + I'(\nu_0)(\nu-\nu_0) + o(\nu-\nu_0) \\
       &= I(\nu_0) + A\nu_0^{-(1+\alpha)}(\nu-\nu_0) + o(\nu-\nu_0) \\
       &= C_1\nu + C_2 + o(\nu-\nu_0)
\end{aligned}
\end{equation}
where $\nu_0=800$ MHz, both $C_1$ and $C_2$ are constants independent of $\nu$, and $o(\nu-\nu_0)$ is the high order small amount.

In an intensity mapping survey, many radiation sources may be included in one pointing direction. 
Therefore, the total flux that a telescope receives in a pointing direction can be expressed as 

\begin{equation}
\begin{aligned}
    f &= \sum_{i}^{N}C_i\nu+\sum_{i}^{N}D_i+\sum_{i}^{N}o(\nu-\nu_0)\\
      &= C\nu + D + o(\nu-\nu_0) \\
      &\approx C\nu + D
\end{aligned}
\label{equ_flux}
\end{equation}
where $N$ is the total number of sources emitting synchrotron or bremsstrahlung in the pointing direction, both $C$ and $D$ are constants independent of $\nu$.
As shown in Equation \ref{equ_flux}, the summation of synchrotron and bremsstrahlung radiation in each pointing direction of a telescope can be approximately modeled as a one-order polynomial.

\end{document}